\documentclass[journal=ancac3,manuscript=article,layout=twocolumn]{achemso}
\setkeys{acs}{articletitle = true}
\setkeys{acs}{email = true}
\usepackage[version=3]{mhchem} 
\usepackage{graphicx}
\usepackage{natmove} 

\author{Ebrahim Forati}
\author{George W. Hanson}
\email{george@uwm.edu}
\affiliation[Department of Electrical Engineering, University of Wisconsin-Milwaukee, 3200 N. Cramer St., Milwaukee, Wisconsin 53211, USA]{University of Wisconsin-Milwaukee}

\author{Alexander B. Yakovlev}
\affiliation[Center for Applied Electromagnetic Systems Research (CAESR), Department of Electrical Engineering, University of Mississippi, University, Mississippi 38677-1848, USA]{University of Mississippi}
\author{Andrea Al\`{u}}
\affiliation[Department of Electrical and Computer Engineering, University of Texas at Austin, Austin, Texas 78712, USA]{The University of Texas at Austin}

\title[planar hyperlens]
  {A planar hyperlens based on a modulated graphene monolayer}

\abbreviations{SPP,THz}

\makeatother

\begin{document}

\begin{abstract}
The canalization of terahertz surface plasmon polaritons using a modulated graphene monolayer 
 is investigated  for subwavelength imaging. An anisotropic surface conductivity formed
by a set of parallel nanoribbons with alternating positive and negative
imaginary conductivities is used to realize the canalization regime required for hyperlensing.
The ribbons are narrow compared to the wavelength, and are created electronically by gating a graphene layer over a corrugated ground plane. Good quality canalization of surface plasmon polaritons is shown in the terahertz even in the presence of realistic  loss  in graphene, with relevant implications for subwavelength imaging applications.
\end{abstract}


\section{========================\label{sec:Introduction} }

Graphene, the first 2D material to be practically realized\cite{Novoselov},
has attracted great interest in the last decade. The
fact that electrons in graphene behave as massless Dirac-Fermions
leads to a variety of anomalous properties \cite{Castro,Luo},
such as charge carriers with ultra high-mobility and long mean-free
paths, gate-tunable carrier densities, and anomalous quantum Hall
effects \cite{Zhang}. Graphene's electrical properties have been
studied in many previous works \cite{Falkovsky,Falkovsky-1,Mikhailov,PGusynin,Gusynin-1,Peres,W_Hanson,Hanson_1,Peres_1,Ziegler}
and are often represented  by a local complex surface conductivity
given by the Kubo formula \cite{Ashcroft,Gusynin}. Since its surface
conductivity leads to attractive surface plasmon properties, graphene
has become a good candidate for plasmonic applications,
especially in the terahertz (THz) regime \cite{Vakil,Johan_Christensen,GWHanso,Nikitin,Sounas,Apell,Forati}.

Surface plasmons (SPs) are the collective charge oscillations at the
surface of plasmonic materials. SPs coupled with photons form the
composite quasi-particles known as surface plasmon polaritons (SPPs).
Theoretically, the dispersion relationship for SPPs on a surface can
be obtained as a solution of Maxwell's equations \cite{Raether}.
In this approach it is easy to show that, in order to support the
SPP, 3D materials with negative bulk permittivities ($e.g.,$ noble metals)
or 2D materials with non-zero imaginary surface conductivities ($e.g.,$ graphene) are essential. Although SPPs on metals and on graphene have
considerable qualitative similarities, graphene SPPs generally
exhibit stronger confinement to the surface, efficient wave localization up to mid-infrared frequencies \cite{Monero,GWHanso}, and they
are highly  tunable (which is one of their most unique and important properties)\cite{Luo}.
Applications of graphene SPPs include electronics \cite{Berger,Mueller,Nair},
optics \cite{Bonaccorso,Xia,Mak}, THz technology \cite{Otsuji,Docherty,Carrier},
light harvesting \cite{Jiang}, metamaterials \cite{Chen}, and medical
sciences \cite{Dong,Szunerits}. 

In this work we study the canalization
of SPPs on  graphene, which can have direct applications  for sub-wavelength
imaging using THz sources.

Sub-wavelength imaging using metamaterials was first reported by Pendry in 2000 \cite{Pendry}.
His technique \cite{Veselago} was based on backward waves, negative refraction and amplification
of evanescent waves. More recently, another more robust venue for subwavelength imaging was proposed, based on metamaterials operating in the so called ``canalization regime'' \cite{Ikonen,Narimanov,Alessandro}. In this case, the structure (acting as a transmission medium)
transfers sub-wavelength images from a source plane to an image plane over distances of several wavelengths, without diffraction \cite{Belov}.
This form of super-resolving imaging, or hyperlensing, can also be realized by a uniaxial wire medium \cite{Belov_WM}. In these  schemes, all spatial harmonics (evanescent and propagating) propagate with the same phase velocity  from the near- to the far-field. In this paper we discuss the canalization of SPPs on a modulated graphene monolayer. In Ref. \cite{mafi}, it was shown that the near field of a vertical point source placed in close proximity to a graphene monolayer couples primarily to the field of an SPP strongly confined to the monolayer. By creating an anisotropic graphene surface as alternating graphene nanoribbons with positive and negative imaginary surface conductivities, we achieve  SPP canalization and hyperlensing of the near-field of an arbitrary source.

To achieve canalization, it is necessary to realize a flat isofrequency
contour\cite{Belove_simovski}. Here, taking the same definition for canalization as for a 3D material,
we first study the conditions for canalization of SPPs on
a 2D material such as graphene. Then, a practical geometry is proposed
and verified for the hyperlens implementation.

\section{Theory and results}

\begin{figure}[t!]
\begin{centering}
\includegraphics[width=3.5in]{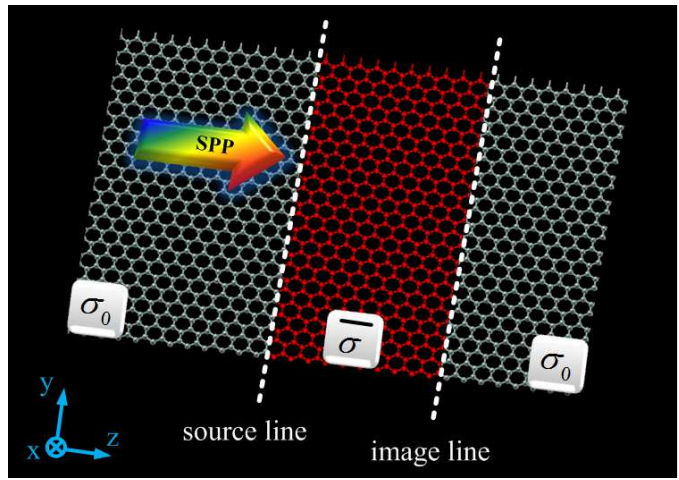} 
\par\end{centering}

\caption{An infinite graphene layer in the $yz-$plane. The conductivity of
 graphene is isotropic ($\sigma_{0}$) everywhere except in the
red region, where it is anisotropic ($\overline{\sigma}$). The
anisotropic region will be created by a suitable gate bias.}
\label{fgr:geometry}
\end{figure}

Figure \ref{fgr:geometry} shows an infinite graphene layer in the $yz-$plane suspended
in vacuum. Its surface conductivity is assumed  isotropic ($\sigma_{0}$) everywhere except in the region between the source
and image lines (red colored region), which is anisotropic and is given as

\begin{equation}
\overline{\sigma}=\sigma_{y}\hat{\mathbf{y}}\hat{\mathbf{y}}+\sigma_{z}\hat{\mathbf{z}}\hat{\mathbf{z}}=-j\left(\sigma_{y}^{i}\hat{\mathbf{y}}\hat{\mathbf{y}}+\sigma_{z}^{i}\hat{\mathbf{z}}\hat{\mathbf{z}}\right),\label{eq:sigma}
\end{equation}
where for now $\sigma_{y,z}$ are assumed to be imaginary-valued, (to be generalized later)
$\sigma_{y,z}=0-j\sigma_{y,z}^{i}$. For an SPP traveling over such an anisotropic
graphene layer, it is possible [see Supporting Information (SI)] to show that
the governing dispersion relation is

\begin{equation}
k_{z}^{2}\left(\frac{\sigma_{z}^{i}}{\sigma_{z}^{i}+\sigma_{y}^{i}}\right)+k_{y}^{2}\left(\frac{\sigma_{y}^{i}}{\sigma_{z}^{i}+\sigma_{y}^{i}}\right)-k_{0}^{2}=\qquad\qquad\label{eq:isocontour-1}
\end{equation}
\[
\qquad\frac{k_{0}k_{x}}{\sigma_{z}^{i}+\sigma_{y}^{i}}\left(\frac{2}{\eta_{0}}-\frac{\eta_{0}\sigma_{y}^{i}\sigma_{z}^{i}}{2}\right),
\]
where $k_{0}$ is the wavenumber in free space,  $\eta_{0}=\sqrt{\mu_{0}/\varepsilon_{0}}$ is the intrinsic
impedance of vacuum, $k_{x}=\sqrt{k_{y}^{2}+k_{z}^{2}-k_{0}^{2}}$, and the 2D spatial Fourier transform variables are $\left(y,z\right)=\left(k_{y},k_{z}\right)$.

From (\ref{eq:isocontour-1}), an ideal canalization regime
can be realized when 
\begin{equation}
\sigma_{y}^{i}\rightarrow0;\quad\sigma_{z}^{i}\rightarrow\infty,\label{eq:cond}
\end{equation}
 simultaneously, such that (\ref{eq:isocontour-1}) becomes

\begin{equation}
k_{z}=k_{0},\label{eq:canalization}
\end{equation}

\noindent independent of $k_{y}$. Equation (\ref{eq:canalization}) implies that all of the transverse
spatial harmonics ($k_{y}$ of the SPPs) will propagate with the same wavenumber
(phase velocity) in the $z$-direction. In this situation, which is analogous to the canalization regime in 3D metamaterials, any SPP distribution at the source line in
Fig. \ref{fgr:geometry} will be transferred to the image line without diffraction or any phase distortion.
Condition (\ref{eq:cond}) is somewhat  analogous to the condition required
for canalization of 3D waves in Ref. \cite{Ramakrishna}, but with the difference that here the extreme parameters (\ref{eq:cond}) yield a finite wave number, equal to the background medium surrounding the modulated graphene layer, and not zero as for the 3D case. This is to be expected, since the canalized SPPs still need to be above the light cone to avoid radiation and leakage in the background medium. Quite peculiarly, it follows from (\ref{eq:canalization}) that the confinement in the transverse ($x$) direction of each SPP is proportional to its spatial frequency along $y$, i.e., $k_{x} = k_{y}$.

It might seem difficult to find a natural 2D material providing (\ref{eq:cond}) for canalization. However,
it can be shown [see SI] that a modulated isotropic conductivity
$\sigma\left(z\right)$ can act as an effective anisotropic conductivity,

\begin{equation}
\sigma_{y}^{\mathrm{eff}}=\frac{1}{T}\intop_{\left\langle T\right\rangle }\sigma\left(z\right)dz,\label{eq:eff_1}
\end{equation}

\begin{equation}
\frac{1}{\sigma_{z}^{\mathrm{eff}}}=\frac{1}{T}\intop_{\left\langle T\right\rangle }\frac{1}{\sigma\left(z\right)}dz,\label{eq:eff_2}
\end{equation}
where $\sigma\left(z\right)$ is assumed to be periodic with period
$T$, and the integrations are over one period. Note that $T$ should
be small compared to the wavelength in order to provide valid effective parameters.
Therefore, if the isotropic conductivity of graphene is properly  modulated ($e.g.,$ by electrical gating or chemical doping), its effective
anisotropic conductivity can indeed satisfy (\ref{eq:cond}).

In the following, two conductivity modulations will be analyzed
whose effective anisotropic conductivities satisfy (\ref{eq:cond})
and are thus in principle  capable of canalizing SPPs. Since we will use full-wave
simulations to confirm the canalization geometries, a section in the SI
is dedicated to modeling of graphene in commercial simulation codes
using finite-thickness dielectrics.

\subsection{Modulated conductivity using ridged ground planes}

\begin{figure}[!t]
 \includegraphics[width=3.5in]{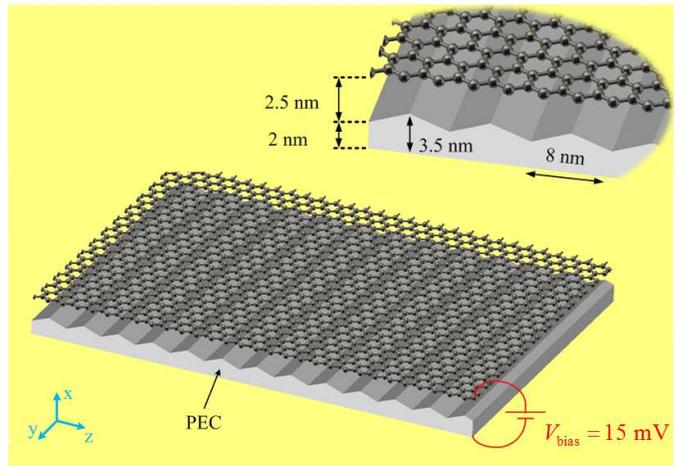}
\caption{Triangular ridged ground plane for achieving conductivity modulation (leading to a soft-boundary profile).}
\label{fgr:triangular}
\end{figure}

\begin{figure}[!tbh]

\includegraphics[width=3.5in]{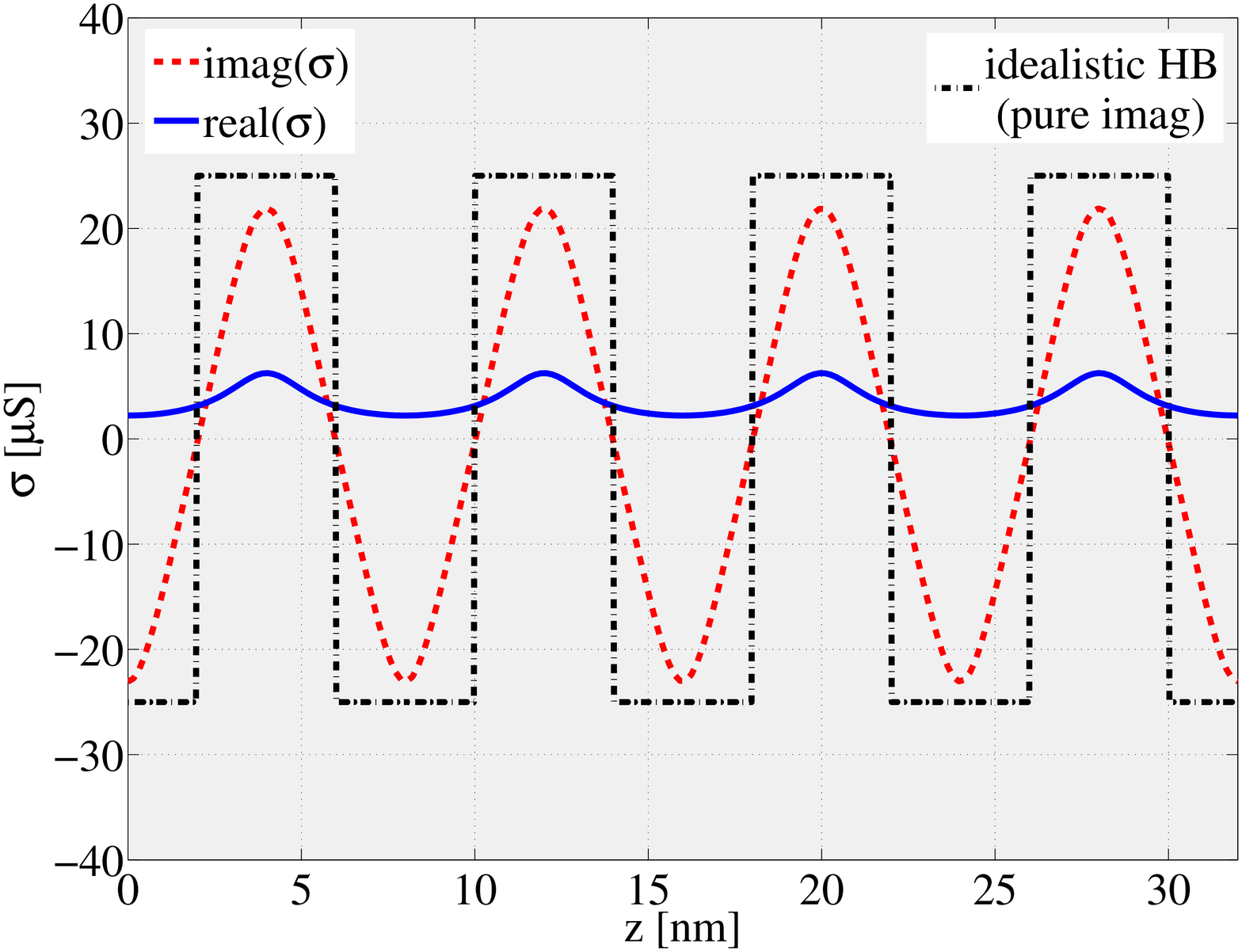} 

\caption{The conductivity distributions  resulting from the bias modulation scheme depicted in Fig. \ref{fgr:triangular}. Also shown is the idealized hard-boundary case discussed in the SI.}
\label{fgr:triangular_cond}
\end{figure}

In previous canalization metamaterials, or hyperlenses, using alternating positive and negative dielectrics, an idealized, abrupt transition has been assumed between layers. For graphene, this would be analogous to strips having abrupt transitions between positive and negative imaginary-part conductivities. We refer to this as the hard-boundary case, and analyze it in detail in the SI. However, given the finite quantum capacitance of graphene, such an abrupt transition is impossible to achieve. A more realistic modulation  scenario for a conductivity profile  satisfying (\ref{eq:cond}) can be obtained in the geometry of Fig. \ref{fgr:triangular}. It consists of an infinite sheet of graphene gated
by a ridged ground plane, as shown in the insert of Fig. \ref{fgr:triangular}. Performing a 
static analysis, it is possible to obtain the charge density on the graphene layer, which may in turn provide the chemical potential and the conductivity of graphene following a method analogous to Ref. \cite{Forati}. Figure \ref{fgr:triangular_cond} shows the calculated conductivity of the graphene layer as a function of $z$ (using the complex conductivity predicted by the Kubo formula; see Ref. \cite{W_Hanson} for the explicit expression for $f=10\,\mathrm{THz}$, $T=3\,\mathrm{K}$, $\Gamma=0.215\,\mathrm{meV}$).

Two important conclusions can be drawn from Fig. \ref{fgr:triangular_cond}: i) the
imaginary part of conductivity dominates the real part, as desired, and ii) its distribution is almost perfectly sinusoidal, which, after insertion into (\ref{eq:eff_1}) and (\ref{eq:eff_2}), satisfies (\ref{eq:cond}). Therefore, the geometry  of Fig. \ref{fgr:triangular} may be expected to support canalization. The resulting graphene nanoribbons have a realistic smooth variation in conductivity; we
refer to this geometry as the soft-boundary scenario, considered in the following.

\begin{figure}[!t]
\begin{minipage}[t]{0.45\textwidth}%
\begin{center}
\includegraphics[width=3.5in]{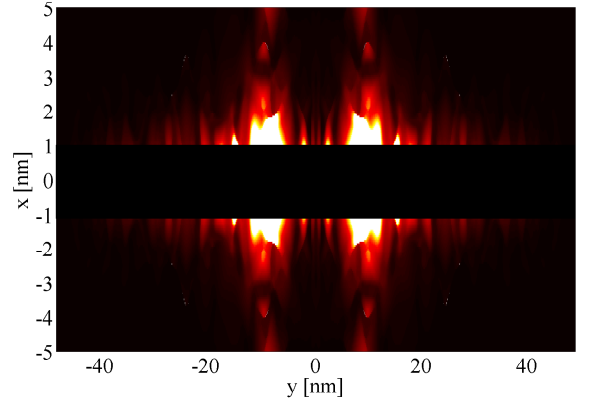} 
\par\end{center}%
\end{minipage}\hfill{}%
\begin{minipage}[t]{0.45\textwidth}%
\begin{center}
\includegraphics[width=3.5in]{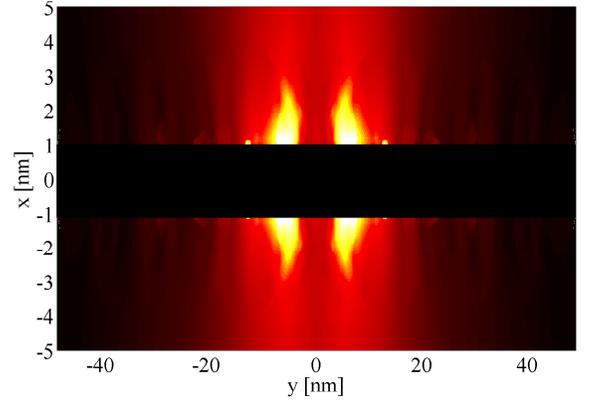} 
\par\end{center}%
\end{minipage}\caption{The normalized $x$-component of the electric field at the source (top)
and image (bottom) planes of the modulated graphene surface. Source and
image lines are at separated by $2\lambda_{\mathrm{SPP}}$ (the region $-1<x<1$ is the dielectric slab model of graphene).}
\label{fgr:SB}
\end{figure}

\begin{figure}[!t]
 \includegraphics[width=3.5in]{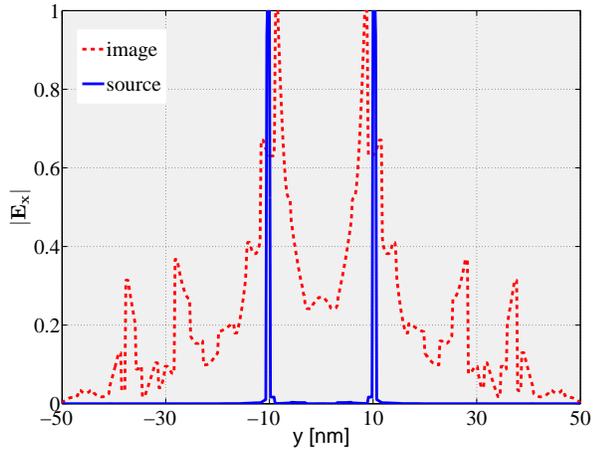}\caption{The normalized $x$-components of the electric field at the source and image lines on the surface of the modulated  graphene ($x=1\, $nm).}
\label{fgr:SB_2D}
\end{figure}

\begin{figure}[!t]
\begin{minipage}[t]{0.45\textwidth}%
\begin{center}
\includegraphics[width=3.5in]{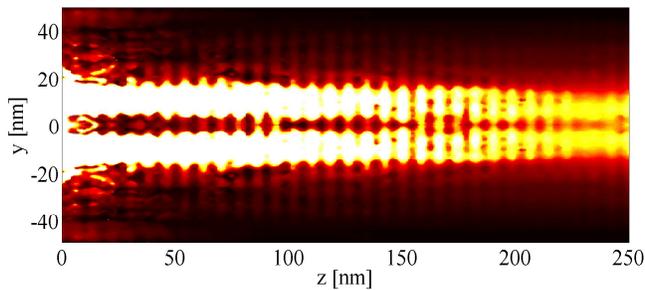} 
\par\end{center}%
\end{minipage}\hfill{}%
\begin{minipage}[t]{0.45\textwidth}%
\begin{center}
\includegraphics[width=3.5in]{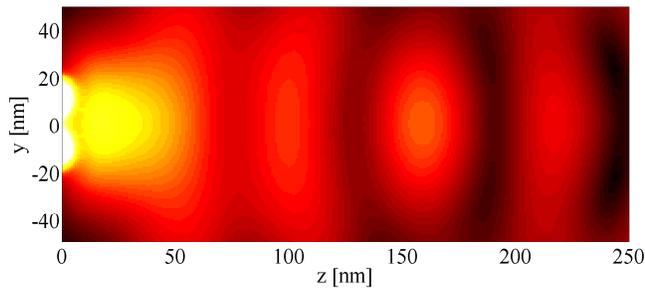} 
\par\end{center}%
\end{minipage}\caption{Normalized $x$-component of the electric field above the modulated graphene surface (top) and a homogenous graphene surface (bottom).}
\label{fgr:SB_above}

\end{figure}

As an example, two point sources are placed in front of the source line in Fig. \ref{fgr:geometry}, exciting  SPPs on the graphene layer. The point sources are separated by $20\,$nm$=0.15\lambda_{\mathrm{SPP}}$, where $\lambda_{\mathrm{SPP}}=133\, $nm using (S.2) in the Supporting Information, and the canalization area (the region between the source and the image lines) has length $2\lambda_{\mathrm{SPP}}=250\, $nm and width of 100$\,$nm (which is large compared to the separation between sources).

Figure \ref{fgr:SB} shows the $x$-component of the electric field at the source line and image line (at the end of the modulated region). The plot of the normalized $x$-component of the field at $x=1\, $nm is shown in Fig. \ref{fgr:SB_2D}, while Figure \ref{fgr:SB_above} shows the $x$-component of the electric field above the modulated graphene surface, and a homogenous graphene surface with conductivity $\sigma=-j23.5\,\mu $S. This shows quite strikingly how  the canalization occuring on the modulated graphene can avoid the usual diffraction expected on a homogeneous layer. Figs. S.3-S.5 in the SI show consistent results  for the hard-boundary case.

\begin{figure}[!t]
 \includegraphics[width=3.5in]{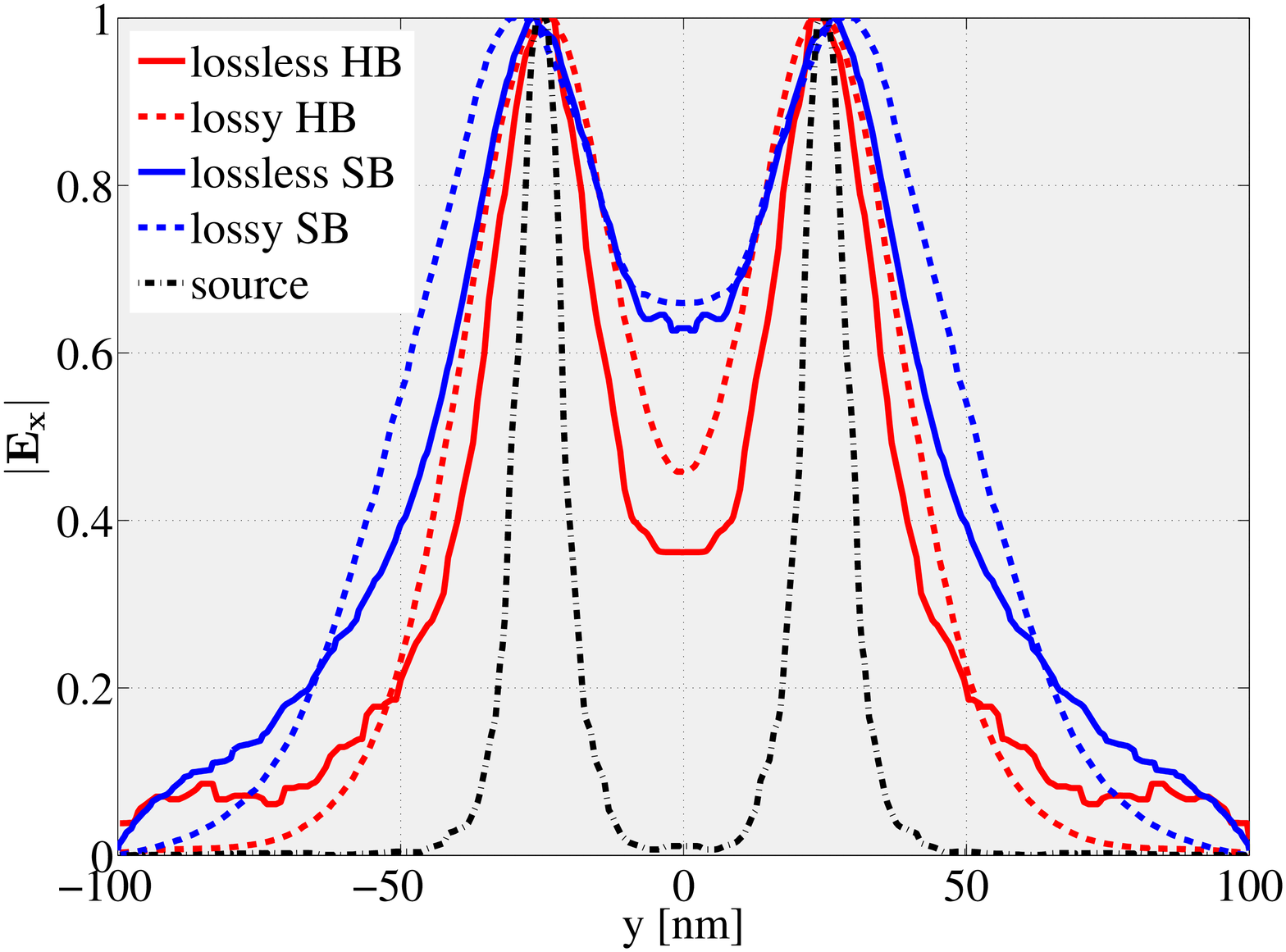}\caption{The effect of loss on the image canalization  for hard- and soft-boundary bias modulations.}
\label{fgr:loss}
\end{figure}

It is easy to show that (\ref{eq:eff_1}) and (\ref{eq:eff_2}) cannot be exactly satisfied if the conductivity includes loss (i.e., the real part of $\sigma$). Therefore as loss  increases,
the phase velocities will  differ among various spatial components and, as a result, one would expect to see a blurred image, and eventually
no image, as loss  further increases. To investigate this deterioration effect, we decrease the canalization length to $200\, $nm$=1.5\lambda_{\mathrm{SPP}}$ and increase the separation between sources to $50\, $nm$=0.4\lambda_{\mathrm{SPP}}$  (which we found necessary to maintain accuracy in the simulation).
The geometry is then simulated for soft- and hard-boundary cases (with and without loss for each case) and the $x$- components of the electric field at $x=10\, $nm are shown in Fig. \ref{fgr:loss}. The curves are calculated in the image line at a distance $1\, $nm above the graphene surface.

Comparison between the four curves in Fig. \ref{fgr:loss} shows that the lossless
hard- and soft-boundary examples yield similar results, as expected since their effective surface conductivity satisfies (\ref{eq:cond}) exactly. In fact, as long as the period is small compared to the wavelength,
any modulation which has half-wave symmetry will satisfy (\ref{eq:cond}), leading to perfect canalization.

However, adding loss causes the effective surface conductivities to
have  non-vanishing real parts, and therefore (\ref{eq:cond}) cannot
be exactly satisfied. In the lossy case, the modulation scheme is
important, since it affects how closely (\ref{eq:cond}) can be satisfied.
For example, Fig. \ref{fgr:loss} shows that the idealistic hard-boundary model exhibits  better resolution than the realistic soft-boundary model.

Image degradation due to loss can be lessened by working at higher frequencies. In fact, the maximum of the ratio Im($\sigma$)/Re($\sigma$) may be increased by adjusting the chemical potential at higher frequencies. In the SI the ratio Im($\sigma$)/Re($\sigma$) is plotted as a function of chemical potential and frequency, and its optimal value for three different frequencies is used to simulate to the same geometry. The simulation results confirm the improvement of canalization as frequency increases.

Our results show that a triangular ridged ground plane to bias the graphene monolayer indeed allows canalization and hyperlensing, since its effective conductivities given by (\ref{eq:eff_1})
and (\ref{eq:eff_2}) satisfy (\ref{eq:cond}). However, there are
many possible $\sigma\left(z\right)$ functions that, after inserting them into (\ref{eq:eff_1}) and (\ref{eq:eff_2}),
will satisfy (\ref{eq:cond}). As an example, the sinusoidal conductivity
of Fig. \ref{fgr:triangular_cond} can also be implemented using a rectangular ridged ground plane (details are shown in the SI).

\section{Conclusions and future scope}

We have analyzed the possibility to produce in-plane canalization of SPPs on a 2D surface, with particular emphasis on its realization in a realistically modulated graphene  monolayer, resulting in a planarized 2D hyperlens on graphene.  We envision the use of this effect on a ridged ground plane for sub-wavelength imaging of THz sources and to arbitrarily  tailor the front wave of an SPP by suitably  designing the boundary of the canalization region. 

\section{Methods}

Simulations were performed with CST Microwave Studio \cite{CST} using a dielectric slab model of graphene, with the conductivity coming from the Kubo formula \cite{Ashcroft,Gusynin}. The Supplemental Information describes in detail the model, contains proofs of various equations appearing in the text, and presents additional results.


\makeatletter 
\renewcommand{\thefigure}{S.\@arabic\c@figure}
\renewcommand{\theequation}{S.\@arabic\c@equation}
\makeatother

\section{Supporting Information}

\subsection{On the modeling of  graphene layer by a thin dielectric}

Modeling graphene as a 2D surface having an appropriate value of surface conductivity $\sigma$  is an accurate approach for a semiclassical analysis (e.g., the Drude model for intraband contributions
has been verified experimentally \cite{Basov,Choi,Choi2}, and the interband model and
the visible-spectrum response have also been verified \cite{Choi2}).  However,
often it is convenient to model graphene as a thin dielectric layer,
which is easily implemented in typical electromagnetic simulation
codes. It is common to consider an equivalent dielectric slab with
the thickness of $d$ and a 3D conductivity of $\sigma_{\mathrm{3D}}=\sigma/d$.
The corresponding bulk (3D) relative permittivity is \cite{Vakil}

\begin{equation}
\varepsilon_{\mathrm{3D}}=1+\frac{\sigma}{j\omega\varepsilon_{0}d},\label{eq:3D}
\end{equation}
where $\omega$ is the angular frequency. However, for calculations
in which the geometry is discretized (e.g., in the finite-element method), fine
features in the geometry such as an electrically-thin slab demand finer discretization,
which in turn requires more computational costs. Thus, whereas sub
$1\,\mathrm{nm}$ thickness values may seem more physically-appropriate, numerical
considerations often lead to the use of a thicker material. As an
example, in Ref. \cite{Vakil} the thickness of the dielectric slab is
assumed to be $1\,\mathrm{nm}$. 

However, the accuracy of the dielectric model degrades as the thickness
of the slab increases. Since this model is widely adopted, yet a detailed
consideration of this effect has not been previously presented, we
briefly consider this topic below.

Consider a transverse magnetic SPP on an infinite
graphene layer. The SPP wavelength using the 2D conductivity is \cite{W_Hanson}

\begin{equation}
\lambda_{\mathrm{SPP}}=\lambda_{0}\left(1-\left(\frac{2}{\eta_{0}\sigma}\right)^{2}\right)^{-0.5},\label{eq:Spp}
\end{equation}
where $\lambda_{0}$ is the wavelength in free space.
On the other hand, in Ref. \cite{Alu_2} it is shown that a dielectric
slab with negative permittivity ambient in a medium with positive
permittivity can support two sets of dielectric modes (even and odd).
The odd modes have the wavelength (assuming vacuum as the ambient
medium)

\begin{equation}
\lambda_{\mathrm{odd}}=2\pi\left(-\frac{2}{d}\mathrm{coth}^{-1}\varepsilon_{\mathrm{3D}}\right)^{-1},\label{eq:odd_graphene}
\end{equation}

\noindent where $\varepsilon_{\mathrm{3D}}$ and $d$ are the dielectric
slab permittivity and thickness, respectively. It is shown in Ref. \cite{Alu_2}
that the odd modes can  exist only if

\begin{equation}
\varepsilon_{\mathrm{3D}}<-1.\label{eq:slab}
\end{equation}
It can also be noticed that the modal field distribution outside of
the slab is similar to that of a SPP on graphene. It is easy to show
that in the limit of $d\rightarrow0$ and using (\ref{eq:3D}), the
dielectric-slab odd mode becomes the graphene SPP mode $\lambda_{\mathrm{odd}}\rightarrow\lambda_{\mathrm{SPP}}$.
 It can be shown that (\ref{eq:odd_graphene}) is a good approximation
for $\lambda_{\mathrm{SPP}}$ only if three conditions are satisfied as {[}see
the next sub-section{]}

\begin{equation}
\frac{d}{\lambda_{\mathrm{SPP}}}\ll1,
\end{equation}

\begin{equation}
\left|\sigma\right|\ll\frac{2}{\eta_{0}},\label{eq:cond_2}
\end{equation}

\begin{equation}
\left|\frac{\sigma}{d}\right|>2\omega\varepsilon_{0}.\label{eq:ratio}
\end{equation}

\begin{figure}[tbh]
\begin{centering}
\includegraphics[width=3.5in]{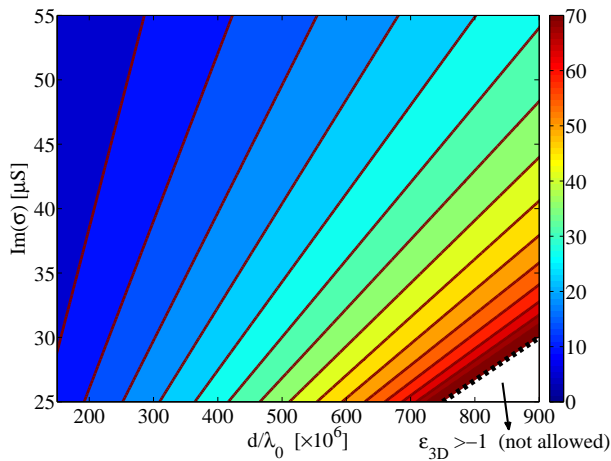} 
\par\end{centering}

\caption{The error (\ref{eq:error}) as a function of the normalized dielectric
thickness and conductivity of graphene. The graph is frequency
independent.}
\label{fgr:slab}
\end{figure}

Equation (\ref{eq:ratio}) is in fact the direct insertion of (\ref{eq:3D})
into (\ref{eq:slab}). Based on (\ref{eq:ratio}), as the $\sigma/d$
ratio increases, the dielectric slab becomes a better approximation
(as long as (\ref{eq:cond_2}) is not violated). To consider this, Fig. \ref{fgr:slab} shows the frequency independent error (\%) of using the dielectric
slab model for graphene as a function of the normalized $d$ and $\sigma$
(assuming $\sigma$ is imaginary-valued),

\begin{equation}
\mathrm{error}(\%)=\frac{\lambda_{\mathrm{odd}}-\lambda_{\mathrm{SPP}}}{\lambda_{\mathrm{SPP}}}\times100.\label{eq:error}
\end{equation}

As a numerical example (using equations (3) and (4) in Ref. \cite{W_Hanson}), for
$d=2 \,$nm, the scattering rate  $\Gamma=0.215\,\mathrm{meV}$, and chemical
potential $\mu_{c}=0.03\,\mathrm{eV}$ at $f=10\,\mathrm{THz}$ and very low temperature ($T=3\,\mathrm{K}$),
the normalized thickness and conductivity will be $d/\lambda_{0}=66.7\times10^{-6}$
and $\sigma=1.1-j23\,\mu$S which leads to an error of $4.9\%$.
This is set as the maximum error that is allowed in the rest of this
work.

\subsection{Proof of (\ref{eq:ratio})}

From (\ref{eq:odd_graphene}),

\begin{equation}
\mathrm{coth}\left(\frac{d\left|\beta_{\mathrm{odd}}\right|}{2}\right)=\frac{\sigma^{i}}{\omega\varepsilon_{0}d}-1\label{eq:app3}
\end{equation}
 where $\beta_{\mathrm{odd}}=2\pi/\lambda_{\mathrm{odd}}$ and $\sigma=-j\sigma^{i}$.

Assuming $d/\lambda_{\mathrm{odd}}\ll1$, (\ref{eq:app3}) leads to

\begin{equation}
\frac{2}{d\left|\beta_{\mathrm{odd}}\right|}+\frac{d\left|\beta_{\mathrm{odd}}\right|}{6}-...=\frac{\sigma^{i}}{\omega\varepsilon_{0}d}-1.\label{eq:app31}
\end{equation}
 After keeping only the first term of the series in (\ref{eq:app31})
and using the assumption $d/\lambda_{\mathrm{odd}}\ll1$ ,

\begin{equation}
\frac{\left|\lambda_{\mathrm{odd}}\right|}{\lambda_{0}}=\frac{\sigma^{i}\eta_{0}}{2}.\label{eq:app32}
\end{equation}

Comparing (\ref{eq:app32}) and (\ref{eq:Spp}), $\lambda_{\mathrm{odd}}$
is a good approximation of $\lambda_{\mathrm{SPP}}$ only if

\begin{equation}
\left|\sigma^{i}\right|\ll\frac{2}{\eta_{0}}.
\end{equation}

\subsection{Proof of (2)}

For the anisotropic region of Fig. 1, consider a general magnetic
field in the Fourier transform domain as

\begin{equation}
\mathbf{H}=e^{-jk_{y}y-jk_{z}z}\times\qquad\qquad\qquad\qquad\qquad\label{general}
\end{equation}
\[
\left\{ \begin{array}{cc}
\left(H_{x}^{+}\hat{\mathbf{x}}+H_{y}^{+}\hat{\mathbf{y}}+H_{z}^{+}\hat{\mathbf{z}}\right)e^{-\sqrt{k_{y}^{2}+k_{z}^{2}-k_{0}^{2}}x} & x>0\\
\left(H_{x}^{-}\hat{\mathbf{x}}+H_{y}^{-}\hat{\mathbf{y}}+H_{z}^{-}\hat{\mathbf{z}}\right)e^{\sqrt{k_{y}^{2}+k_{z}^{2}-k_{0}^{2}}x} & x<0
\end{array}\right.
\]
 where $H_{x,y,z}^{+,-}$ are constants. Equation (\ref{general})
is chosen so that it satisfies the Helmholtz equation and has the
form of a plasmonic wave.

Using Ampere's law to find the electric field in each region and satisfying
the boundary conditions

\begin{equation}
H_{y}^{+}-H_{y}^{-}=\sigma_{z}E_{z},\label{eq:2-1-3}
\end{equation}

\begin{equation}
H_{z}^{+}-H_{z}^{-}=-\sigma_{y}E_{y},\label{eq:2-1-1-2}
\end{equation}

\begin{equation}
H_{x}^{+}=H_{x}^{-},
\end{equation}

\noindent it is straightforward to show that

\begin{equation}
H_{y}^{-}=-H_{y}^{+},
\end{equation}
 
\begin{equation}
H_{z}^{-}=-H_{z}^{+},
\end{equation}

\begin{equation}
\left[\begin{array}{ccc}
\sigma_{z}jk_{y} & Y & 0\\
jk_{z}\sigma_{y} & 0 & Z\\
jk_{x} & j\omega\varepsilon_{0}k_{y} & j\omega\varepsilon_{0}k_{z}
\end{array}\right]\left[\begin{array}{c}
H_{x}^{+}\\
H_{y}^{+}\\
H_{z}^{+}
\end{array}\right]=0,
\end{equation}

\noindent where $Y=-2j\omega\varepsilon_{0}-\sigma_{z}k_{x}$,
and $Z=-2j\omega\varepsilon_{0}-k_{x}\sigma_{y}.$ Setting the determinant of the above matrix to zero leads to (2).

It is easy to show that in the isotropic limit ($\sigma_{y}=\sigma_{z}=\sigma_{0}$),
(2) simplifies to the well-known dispersion equations
\cite{Mikhailov,W_Hanson} $k_{x}=-\frac{2jk_{0}}{\eta_{0}\sigma_{0}}$, and $k_{x}=-\frac{jk_{0}\eta_{0}\sigma_{0}}{2}$, for transverse magnetic (TM) and transverse electric (TE) surface waves, respectively.
The solution of (2) will lead to a solution
for the SPP with the magnetic field 

\begin{equation}
\mathbf{H}=e^{-k_{x}x-jk_{y}y-jk_{z}z}\times\qquad\qquad\label{eq:magnetiic}
\end{equation}
\[
\left(\hat{\mathbf{x}}+\frac{j\sigma_{z}k_{y}}{2j\omega\varepsilon_{0}+k_{x}\sigma_{z}}\hat{\mathbf{y}}+\frac{j\sigma_{y}k_{z}}{2j\omega\varepsilon_{0}+k_{x}\sigma_{y}}\hat{\mathbf{z}}\right).
\]

In the canalization regime, the SPP given by
(\ref{eq:magnetiic}) is a TM mode with respect to the canalization
direction ($z$-direction in our notation) and its magnetic field has a peculiar 
circular polarization,

\begin{equation}
\mathbf{H}=\left(\hat{\mathbf{x}}+j\hat{\mathbf{y}}\right)e^{-k_{y}\left(x+jy\right)-jk_{0}z}.
\end{equation}

It is also interesting that the confinement in the $x$-direction of each SPP harmonic is proportional to $k_{y}$.

\subsection{Proof of (5) and (6)}

\begin{figure}[!t]
 \includegraphics[width=3.5in]{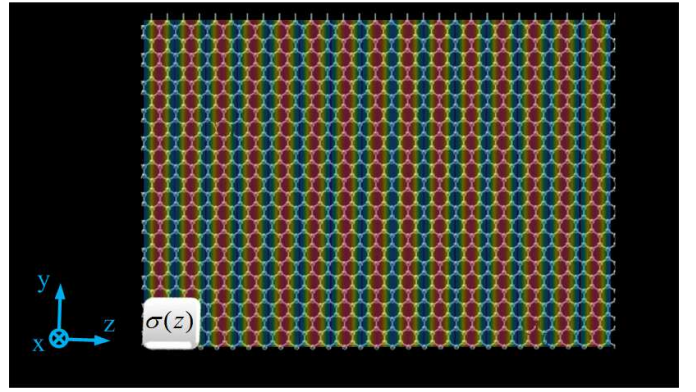}
\caption{An infinite graphene layer with isotropic periodic conductivity of
$\sigma(z)$.}
\label{fgr:effect}
\end{figure}

Assume a sheet of graphene with a periodic isotropic conductivity
in the $z$-direction ($\sigma\left(z\right)=\sigma\left(z+T\right)$) as
shown in Fig. \ref{fgr:effect}. Enforcing a constant, uniform, and $z$-directed surface
current $\left(J_{z}\right)$ on the graphene induces an electric
field on the graphene as

\begin{equation}
E\left(z\right)=\frac{J_{z}}{\sigma\left(z\right)}.
\end{equation}

Defining average parameters leads to

\begin{equation}
E_{\mathrm{av}}=\frac{J_{z}}{\sigma_{\mathrm{av},z}}=\frac{1}{L}\intop_{\left\langle L\right\rangle }\frac{J_{z}}{\sigma\left(z\right)}dz,
\end{equation}

\begin{equation}
\frac{1}{\sigma_{\mathrm{av},z}}=\frac{1}{L}\intop_{\left\langle L\right\rangle }\frac{1}{\sigma\left(z\right)}dz.
\end{equation}

Enforcing a constant, uniform and $y$-directed electric field ($E_{y}$)
induces a surface current on the graphene as 
\begin{equation}
J_{y}\left(z\right)=\sigma\left(z\right)E_{y}
\end{equation}
which is (5).

Defining average parameters leads to

\begin{equation}
J_{y,\mathrm{av}}\left(z\right)=\sigma_{\mathrm{av},y}E_{y}=\frac{1}{L}\intop_{\left\langle L\right\rangle }\sigma\left(z\right)E_{y}dz,
\end{equation}

\begin{equation}
\sigma_{\mathrm{av},y}=\frac{1}{L}\intop_{\left\langle L\right\rangle }\sigma\left(z\right)dz,
\end{equation}
which is (6).

\subsection{Idealized graphene nanoribbons with hard-boundaries }

An idealization of the modulation scheme discussed in the text would consist of  alternating positive
and negative imaginary conductivities, with each strip terminating in a sharp transition between positive and negative values (see Fig. \ref{fgr:HB_geometry}). We assume that all of the strips have the same width $W=4\, $nm and conductivity
modulus $\left|\sigma\right|=23.5\,\mu $S, which is the conductivity
of a graphene layer for $f=10\,\mathrm{THz}$, $T=3\,\mathrm{K}$, $\Gamma=0.215\,\mathrm{meV}$
 and $\mu_{c}=0.022\,\mathrm{eV}$ or $\mu_{c}=0.03\,\mathrm{eV}$ (for positive and negative $\mathrm{Im}\left(\sigma\right)$, respectively).  The chemical potential is  chosen to minimize the loss at the given frequency. In fact,  the ratio $\mathrm{Im}\left(\sigma\right)/\mathrm{Re}\left(\sigma\right)$ is maximized at this frequency (the ratio is 7 for  $\mu_{c}=0.022$eV).  Since the effect of loss was discussed in the text, here we assume an imaginary-valued  conductivity $\sigma=\pm j23.5\,\mu $S.

We refer to this idealized conductivity profile as the hard-boundary case, because
of the step discontinuity (sharp transition) of the conductivity between neighboring strips. This resembles the geometry in Ref. \cite{Ramakrishna}
for canalization of 3D waves in which there are also hard-boundaries
between dielectric slabs with positive and negative permittivites.

\begin{figure}[!t]
\begin{minipage}[t]{0.45\textwidth}%
\begin{center}
\includegraphics[width=3.5in]{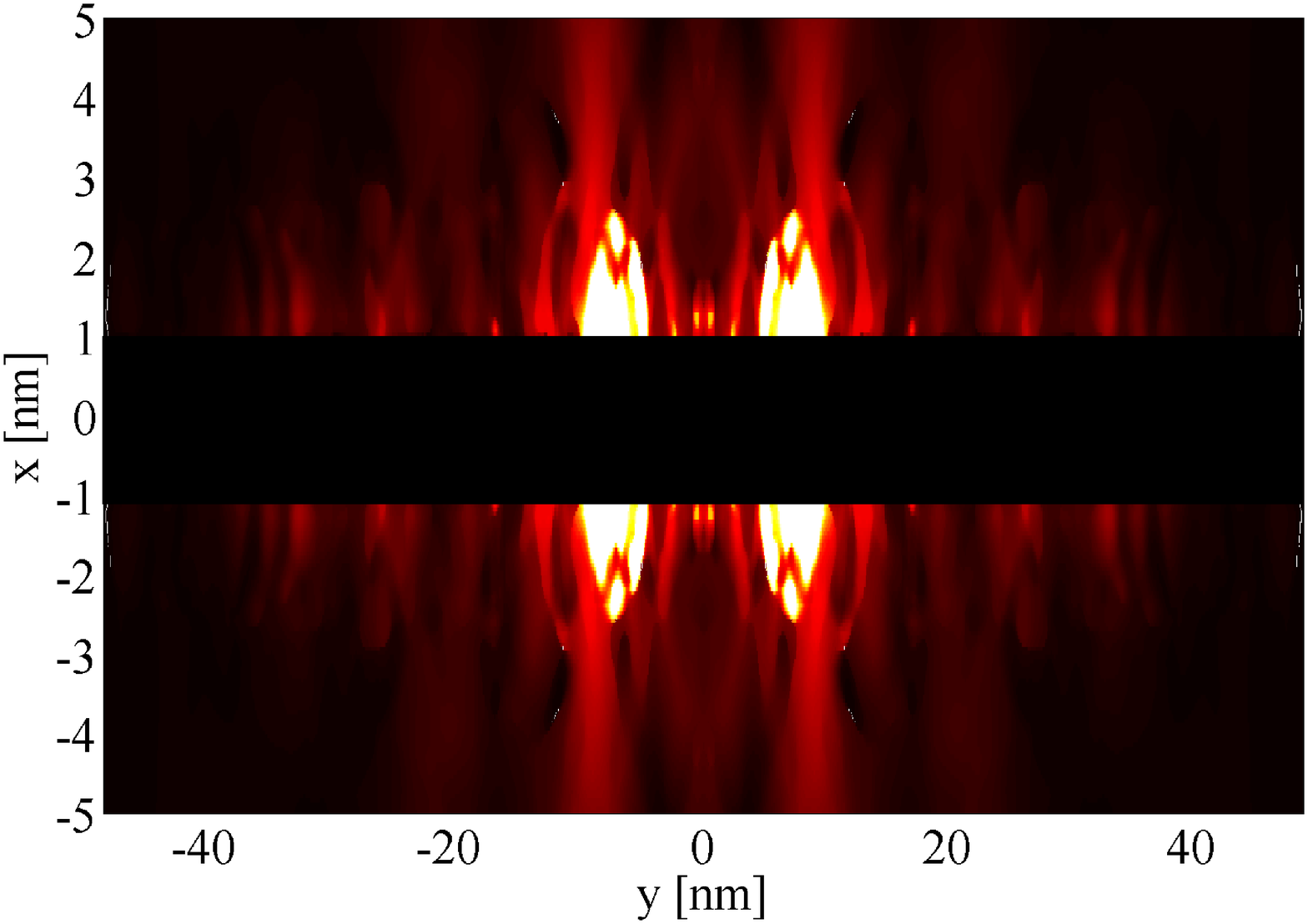} 
\par\end{center}%
\end{minipage}\hfill{}%
\begin{minipage}[t]{0.45\textwidth}%
\begin{center}
\includegraphics[width=3.5in]{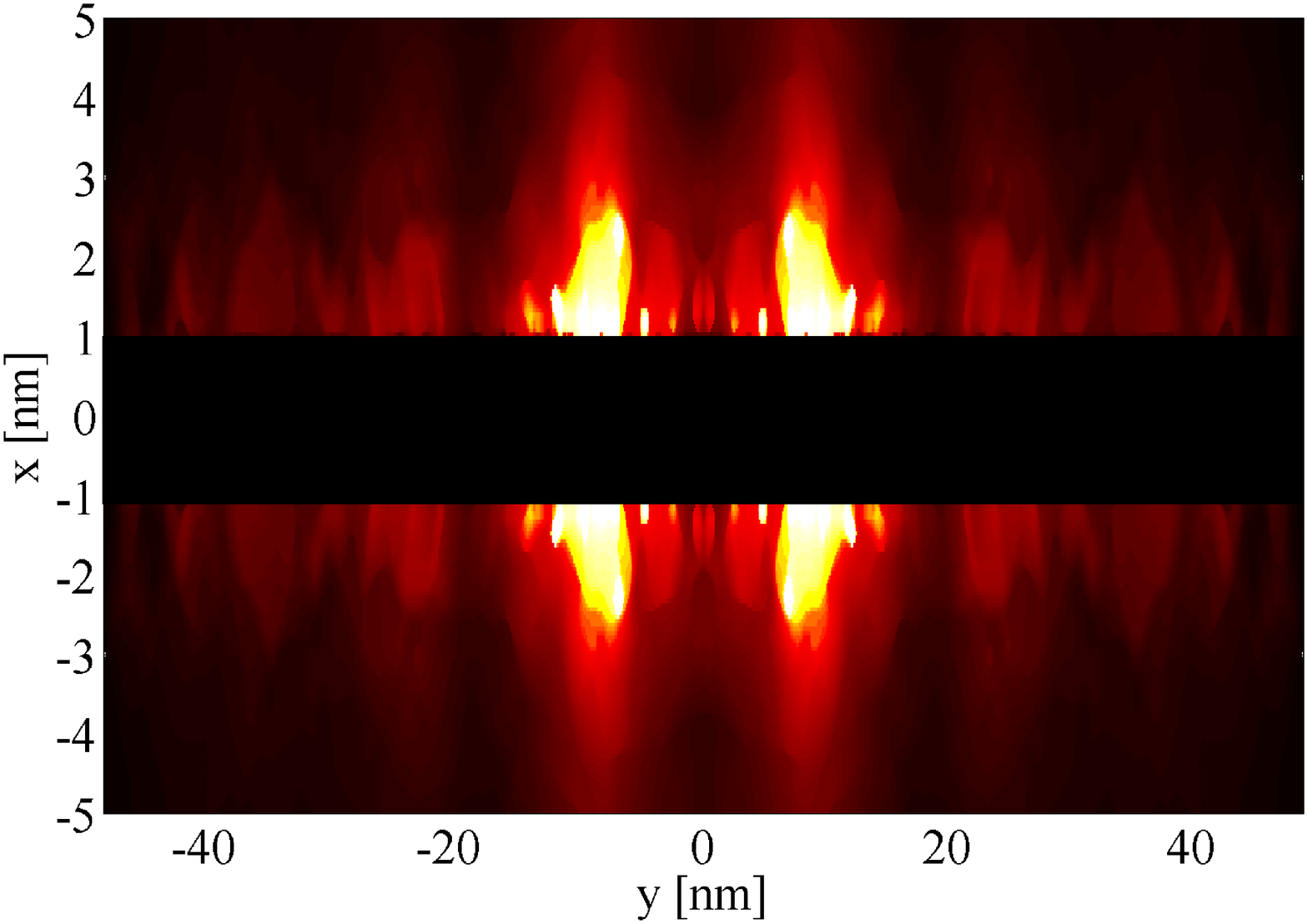} 
\par\end{center}%
\end{minipage}\caption{The normalized $x$-component of the electric field at the source (top)
and image (bottom) planes of the hard-boundary example. Source and
image lines are separated by $2\lambda_{\mathrm{SPP}}$ (the region $-1<x<1$ is the dielectric slab model of graphene).}
\label{fgr:HB}
\end{figure}

\begin{figure}[!tbh]
\includegraphics[width=3.5in]{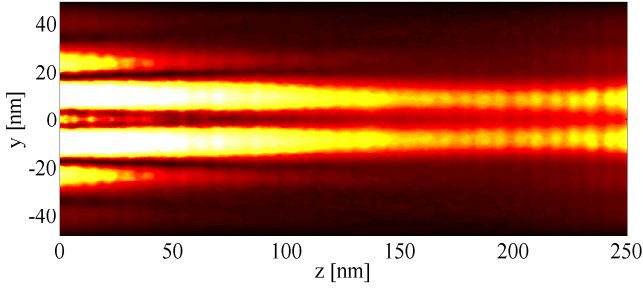}\caption{Normalized $x$-component of the electric field above the graphene surface.}
\label{fgr:HB_above}
\end{figure}

As a simulation example of the hard-boundary case, two point sources are placed in front of the source line in Fig. 1 exciting two SPPs on the graphene layer. The point sources are separated by $20\,$nm$=0.15\lambda_{\mathrm{SPP}}$ where $\lambda_{\mathrm{SPP}}=133\, $nm using (\ref{eq:Spp}), and the canalization area (the region between the source and the image lines) has length $2\lambda_{\mathrm{SPP}}=250\, $nm and width of 100$\,$nm (which is large compared to the separation between sources).
Figure \ref{fgr:HB} shows the normalized $x$-component of the  electric field $\left|E_{x}\right|$
at the source line and image line (at the end of the modulated region). Fig. \ref{fgr:HB_above} shows the normalized $x$-component
of the  electric field above the surface of the graphene ($x=5\, $nm). Note that the region $-1<x<1\, $nm represents the graphene (since we have used a dielectric slab model for graphene with the thickness of $2\, $nm).

\begin{figure}[!t]
 \includegraphics[width=3.5in]{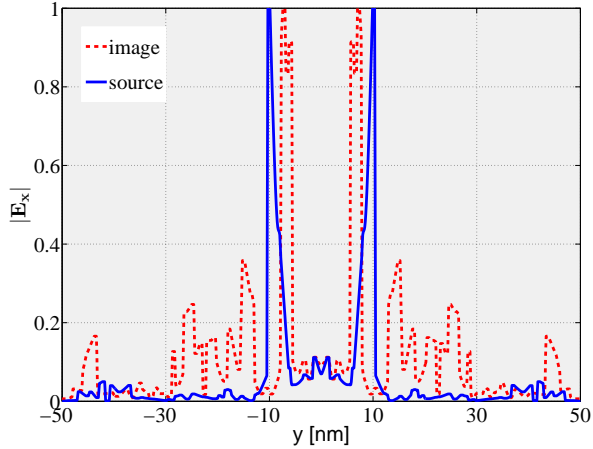}
\caption{The normalized $x$-components of the electric field at the source and
image lines on the surface of the graphene (taken at the height $x=1\, $nm)
for the hard-boundary example.}
\label{fgr:HB_2D}
\end{figure}

Canalization is evident from Figs. \ref{fgr:HB} and \ref{fgr:HB_above}. Figure \ref{fgr:HB_2D} shows the normalized field intensities at the source and image lines just above the graphene surface ($x=1\, $nm). 

\subsection{Simulation setup for the hard- and the soft-boundary examples}

\begin{figure}[!t]
 \includegraphics[width=3.5in]{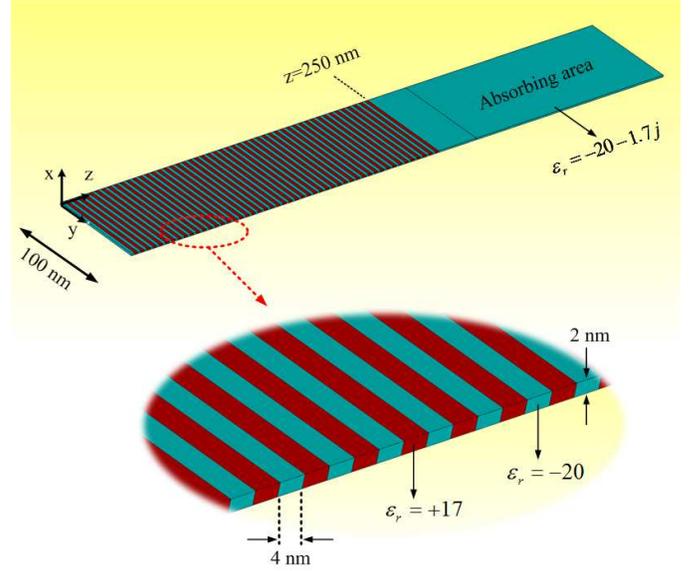}
\caption{The dielectric model of the hard-boundary graphene strip example.}
\label{fgr:HB_geometry}
\end{figure}

Full-wave simulations have been done using CST Microwave Studio \cite{CST}. In this section we consider the dielectric model of graphene. Figure \ref{fgr:HB_geometry} shows the simulation setup of the hard-boundary example. The simulation results are given in Figs. \ref{fgr:HB}-\ref{fgr:HB_2D}. The graphene strips can be modeled with dielectric slabs having thickness
$d=2\, $nm and, using (\ref{eq:3D}), permittivities of $\varepsilon^{-}=-20$ and $\varepsilon^{+}=22$. However, as shown in the insert of Fig. \ref{fgr:HB_geometry}, the permittivity $\varepsilon^{+}=17$ is used rather than $\varepsilon^{+}=22$ because numerical experiments show that that value leads to better canalization. The difference with our analytically-predicted value for best canalization is seemingly because in our analytical model we have disregarded radiation, reflections from discontinuities, and similar effects.

\begin{figure}[!tbh]
\includegraphics[width=3.5in]{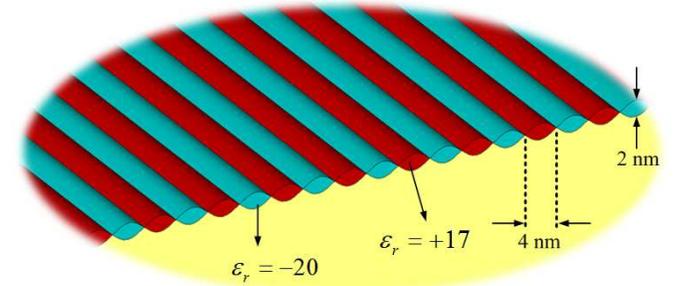}\caption{The dielectric model for the soft-boundary example - constant permitivies
and smoothly-varying thickness model graphene's sinusoidal chemical
potential.}
\label{fgr:SB_geometry}
\end{figure}

For the soft-boundary example, the conductivity of the strips varies smoothly with position. So, applying the dielectric slab model, we could use a dielectric slab with a fixed thickness (e.g., $d=2\, $nm) and a position dependent permittivity given by (\ref{eq:3D}) as 
\begin{equation}
\varepsilon_{\mathrm{3D}}\left(z\right)=1+\frac{\sigma\left(z\right)}{j\omega\varepsilon_{0}d}.
\end{equation}
 However, an alternative method which is easier to implement for simulation
is to consider a dielectric slab with fixed permittivity (or permittivities)
and a position dependent thickness as

\begin{equation}
d\left(z\right)=\frac{\sigma\left(z\right)}{\left(\varepsilon_{\mathrm{3D}}-1\right)j\omega\varepsilon_{0}}.
\end{equation}

Obviously, two different $\varepsilon_{\mathrm{3D}}$ values should be chosen
for different signs of $\sigma\left(z\right)$ so that $d\left(z\right)$
remains positive. This has been done for the conductivity of Fig. 3,
 and the resulting dielectric slab model is shown in Fig. \ref{fgr:SB_geometry}.
Comparison between Fig. \ref{fgr:HB_geometry} and Fig. \ref{fgr:SB_geometry} clearly shows the difference
between the hard- and the soft-boundary examples.


\subsection{The improvement of canalization by increasing the frequency}

\begin{figure}[!t]
 \includegraphics[width=3.5in]{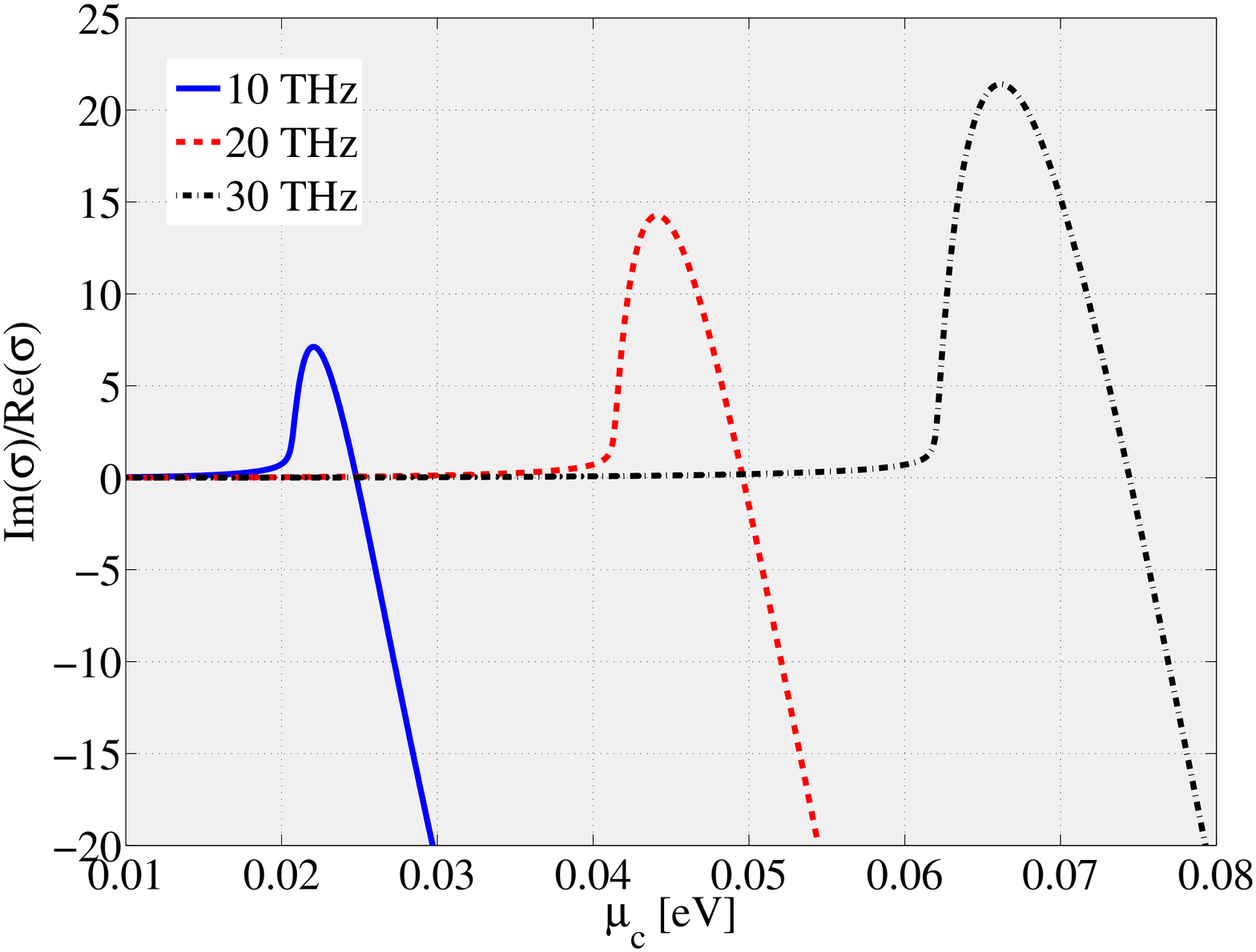}
\caption{The ratio $\mathrm{Im}\left(\sigma\right)/\mathrm{Re}\left(\sigma\right)$ as a function of chemical potential for three different frequencies.}
\label{fgr:ratio}
\end{figure}

\begin{figure}[!t]
\begin{minipage}[t]{0.45\textwidth}%
\begin{center}
\includegraphics[width=3.5in]{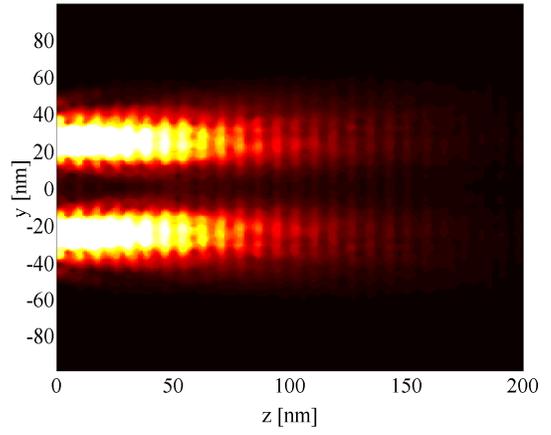} 
\par\end{center}%
\end{minipage}\hfill{}%
\begin{minipage}[t]{0.45\textwidth}%
\begin{center}
\includegraphics[width=3.5in]{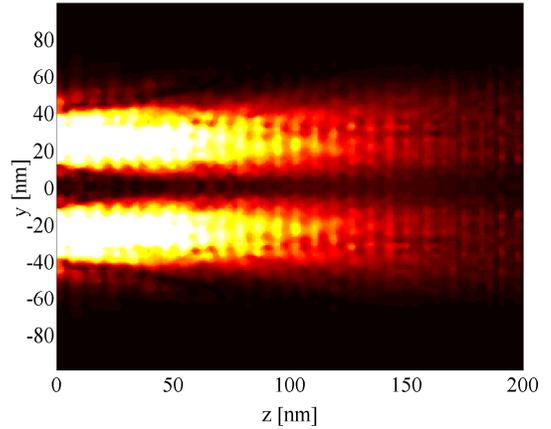} 
\par\end{center}%
\end{minipage}\hfill{}%
\begin{minipage}[t]{0.45\textwidth}%
\begin{center}
\includegraphics[width=3.5in]{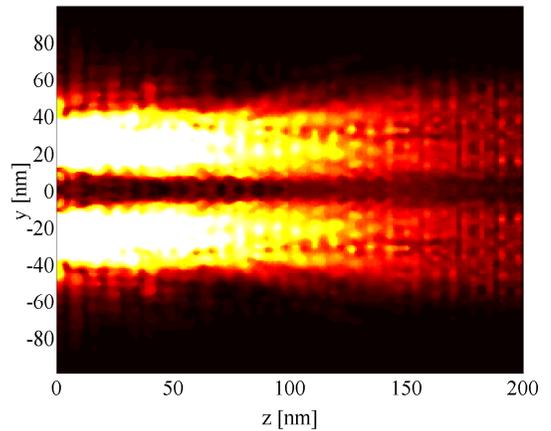} 
\par\end{center}%
\end{minipage}\caption{The normalized $x$-component of the electric field above the graphene surface ($x=2\, $nm) for the peak value of Im($\sigma$)/Re($\sigma$) at 10 THz (top), 20 THz (middle), and 30 THz (bottom). }
\label{fgr:freq}
\end{figure}

Figure \ref{fgr:ratio} shows the ratio Im($\sigma$)/Re($\sigma$) versus chemical potential at three different frequencies, showing that, as frequency increases, loss becomes less important. Note also that the value of chemical potential that maximizes the conductivity ratio is considerably frequency dependent.
In Fig. \ref{fgr:freq} the effect of decreasing loss as a result of the frequency increase is invesigated. To do so, the peak ratio Im($\sigma$)/Re($\sigma$) of the three curves in Fig. \ref{fgr:ratio} are chosen associated with frequencies 10, 20, and 30 THz. These ratios are assigned to a same geometry (and holding frequency constant) and the $x$-component of the electric fields are shown in Fig. \ref{fgr:freq} (the scalings are the same). In this way, all of the electrical lengths (such as the electrical length of the nanoribbons, canalization region, etc.) remain the same and only the effect of loss is incorporated. From Fig. \ref{fgr:freq}, it is obvious that the increase of frequency improves the canalization. However, since the dimensions become smaller, fabrication becomes more difficult.

\subsection{Modulated graphene conductivity using a rectangular ridged ground plane}

\begin{figure}[!tbh]
 \includegraphics[width=3.5in]{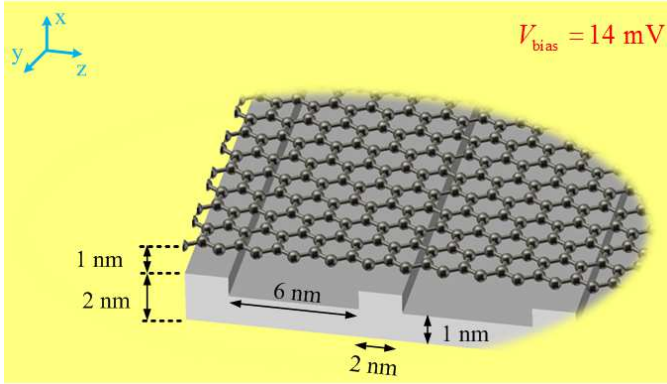}\caption{An alternative geometry with rectangular ridged ground plane to realize the soft-boundary example.}
\label{fgr:rect}
\end{figure}

\begin{figure}[!tbh]
\includegraphics[width=3.5in]{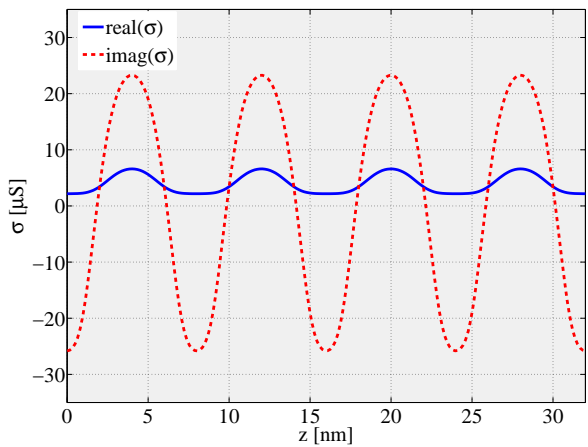} 

\caption{The  conductivity distribution in the geometry of Fig. \ref{fgr:rect}.}
\label{fgr:rect_cond}
\end{figure}

The sinusoidal conductivity of Fig. 3 can be implemented using a rectangular ridged ground plane, as shown in Fig. \ref{fgr:rect}. The conductivity distribution of the geometry in Fig. \ref{fgr:rect} is shown in Fig. \ref{fgr:rect_cond} and is almost identical to  Fig. 3, although their ground plane geometries are different.
Obviously, the ideal canalization behavior of the two geometries is very similar. Interestingly, the rectangular ridged ground  plane has to be non-symmetric (the ratio of groove to ridge is 3) to produce the same conductivity function as the symmetrical triangular ridged ground plane. 

\bibliography{achemso}


\end{document}